# Exploring Nanoscale Photoresponse Mechanisms for Enhanced Photothermoelectric Effects in van der Waals Interfaces

Da Xu[1], Qiushi Liu[1], Boqun Liang[1], Ning Yu[2], Xuezhi Ma[1], Yaodong Xu[1], Takashi Taniguchi[3], Roger K. Lake[1], Ruoxue Yan[2,4], Ming Liu[1,4]

**Addresses:**

[1] Department of Electrical and Computer Engineering, University of California - Riverside, Riverside, California 92521, United States

[2] Department of Chemical and Environmental Engineering, University of California - Riverside, Riverside, California 92521, United States

[3] International Center for Materials Nanoarchitectonics, National Institute for Materials Science, Tsukuba, Ibaraki 305-0044, Japan

[4] Materials Science and Engineering program, University of California - Riverside, Riverside, California 92521, United States

**Abstract:**

Integrated photodetectors are crucial for their high speed, sensitivity, and efficient power consumption. In these devices, photocurrent generation is primarily attributed to the photovoltaic (PV) effect, driven by electron-hole separations, and the photothermoelectric (PTE) effect, which results from temperature gradients via the Seebeck effect. As devices shrink, the overlap of these mechanisms—both dependent on the Fermi level and band structure—complicates their separate evaluation at the nanoscale. This study introduces a novel 3D photocurrent nano-imaging technique specifically designed to distinctly map these mechanisms in a Schottky barrier photodiode featuring a molybdenum disulfide and gold ($MoS_2$-Au) interface. We uncover a significant PTE-dominated region extending several hundred nanometers from the electrode edge, a characteristic facilitated by the weak electrostatic forces typical in 2D materials. Unexpectedly, we find that incorporating hexagonal boron nitride ($h$-BN), known for its high thermal conductivity, markedly enhances the PTE response. This counterintuitive enhancement stems from an optimal overlap between thermal and Seebeck profiles, presenting a new pathway to boost device performance. Our findings highlight the capability of this imaging technique to not only advance optoelectronic applications but also to deepen our understanding of light-matter interactions within low-dimensional systems.

**Main**:

The optoelectronics industry, which significantly influences sectors ranging from high-speed optical communications to LiDAR sensing, is driven by an increasing demand for miniaturized photoreceivers that integrate seamlessly with signal processing chips to deliver superior performance[1,2]. Responding to these technological imperatives, the adoption of ultra-thin materials such as silicon (Si)/germanium (Ge) nanomembranes[3,4] and two-dimensional van der Waals (vdW) materials[5-7], including graphene and transition-metal dichalcogenides, has escalated. These materials are pivotal in minimizing the volume of photoactive material, enhancing operational speed, signal-to-noise ratio, and internal quantum efficiency through optimized photocarrier separation. Moreover, their integration with photonic or plasmonic nanostructures further augments optical absorption and shortens carrier transfer distances, significantly boosting photoresponsivity[8-15].

In nanostructured optoelectronic devices, the PV and PTE effects intricately intertwine, differing from traditional systems[16,17]. In areas like p-n junctions or Schottky contacts, where band structure bending is significant, a strong built-in electric field—produced by rapid spatial variations in charge carrier densities—primarily supports the PV effect. This spatial variation in charge modifies the Seebeck coefficients, which are crucial for the PTE effect. Additionally, localized photon absorption leads to temperature gradients, which are essential for driving the PTE response. Consequently, these temperature gradients, together with variations in the Seebeck coefficient, are central to the PTE effect in these environments[10,18-22]. Thus, PV and PTE effects are not isolated; they coexist and collectively contribute to the overall photocurrent, and they both have complex interplay with the local electronic and thermal environments. Understanding this intricate relationship at the nanoscale is crucial for enhancing the efficiency and sensitivity of next-generation photodetectors, necessitating a detailed exploration of how these effects interact and distribute at the nanoscale.

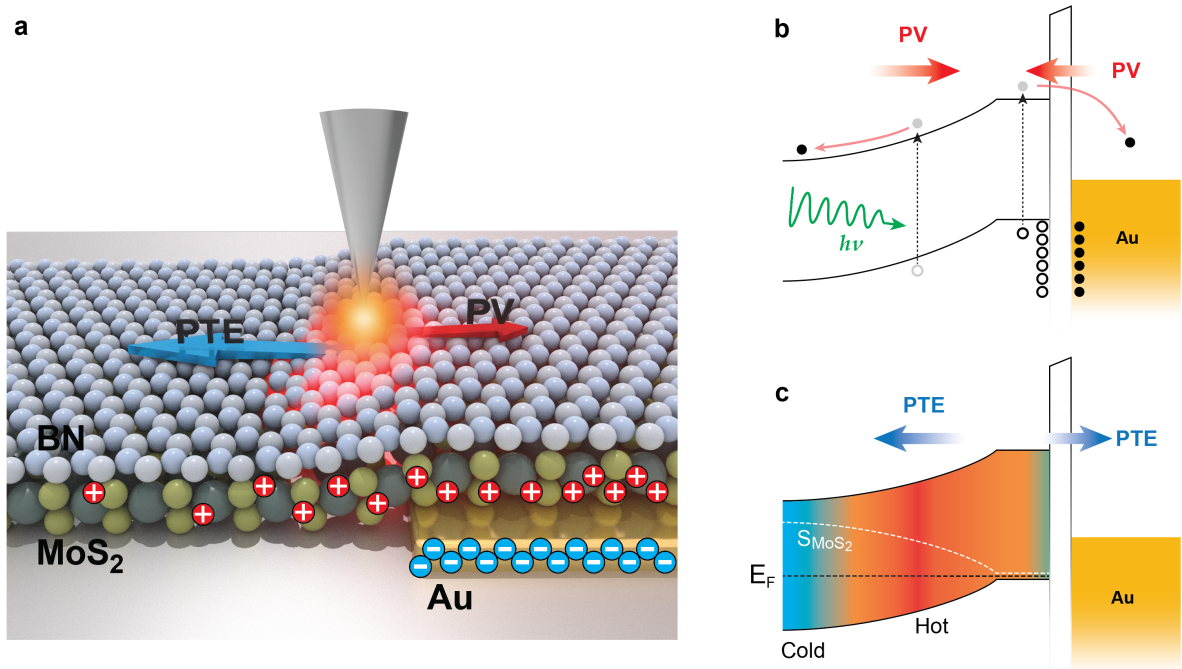

**Figure 1. Nanoscale photocurrent generation mechanisms analysis at a 2D semiconductor – 3D metal contact junction**. **(a)** Schematic of the device, showing a thin MoS2 layer encapsulated by hBN on a gold electrode. **(b)** Energy band diagram under reverse bias highlighting photovoltaic (PV) current generation at the band bending region. **(c)** Illustration of photothermoelectric (PTE) current induced by spatial variations in Seebeck coefficients and temperature gradients across the band bending region.

In this investigation, we deploy 3D scanning near-field photocurrent microscopy to delineate the photocurrent response mechanisms within a biased 2D semiconductor-3D metal Schottky barrier (SB) photodiode at an unprecedented nanoscale resolution. Our 3D imaging methodology, which utilizes the distinctive profiles of approach curves—defined as variations in photocurrent amplitude and polarity across different probe-to-sample distances—enables the independent mapping of PV and PTE distributions. We specifically apply this technique to a $MoS_2$-Au Schottky photodiode. Unlike earlier studies that focused on cross-interface PTE currents from $MoS_2$ to gold within the electrode region, our findings expose a pronounced in-plane PTE-dominated region extending up to a micron from the edge of the gold electrode into the $MoS_2$ channel. This in-plane PTE current, though stronger than the cross-interface PTE effect, has a nanoscale active region and is concealed by the cross-interface PTE effect in conventional far-field photocurrent microscopy. The extensive reach of this PTE region is attributed to the substantial depletion lengths and resulting significant spatial modulation of Seebeck coefficients in 2D materials, driven by their inherently weak electrostatic effects.

As shown in Figure 1, the SB photodiode studied consists of a few-layer $MoS_2$ flake (~10 nm thick) that was mechanically exfoliated and transferred to pre-deposited gold electrodes (30 nm Au / 2 nm Cr) using a capillary force-assisted clean transfer method[23]. This method, akin to the clean electrode-transferring technique[24], prevents the introduction of chemical disorders and defect-induced gap states in the semiconductor that typically result from atom or cluster bombardment and local heating during the metal deposition process. Consequently, the Schottky barrier heights at both the source and the drain approached the Schottky-Mott limit, forming a double SB photodiode. The device was fabricated on a 290-nm $SiO_2$/Si wafer, using the underlying silicon substrate as a back gate to fine-tune the electron doping level in the $MoS_2$ channel, thereby controlling the device's photoresponse characteristics. The $MoS_2$ active layer was strategically positioned atop the electrodes to facilitate near-field optical characterization using a scanning probe. To protect the $MoS_2$ from oxidation and laser ablation, and to enhance thermal management, the entire structure was encapsulated with a thin layer of hexagonal boron nitride (h-BN, ~5 nm).

The 3D photocurrent nano-imaging setup, as shown in Figure 2a, utilizes a superfocused light source, generated through high-external-efficiency nanofocusing at the tip of a tapping-mode atomic force microscope probe[25,26]. This setup meticulously scanned the device surface, with the probe oscillating harmonically above the device. Its rapidly decaying evanescent optical fields excited anharmonic photocurrent signals in the $MoS_2$ gain medium. The resulting photocurrent peaks and troughs, corresponding to the closest and farthest tip-to-sample distances respectively, were demodulated using a multi-order lock-in amplifier to extract the first few harmonic amplitudes $R_n$ and phase $\phi_n$, which were registered at each pixel during the mapping.

It is crucial to recognize that a pair of individual harmonic amplitude $R_n$ and phase $\phi_n$ do not provide a straightforward description of the near-field photocurrent. They are significantly influenced by the shapes of the approach curves, reflecting only the resultant effects of competing photoresponse mechanisms. Consequently, the higher-order demodulation amplitudes $R_n$ and their associated sign functions, denoted as $sgn(\phi_n)$, do not have a direct and unambiguous relationship with the photocurrent excited by the near-field light source.

Instead, the photocurrent is effectively captured by the approach curve $I_{PC}(d)$, reconstructed from all harmonics as $I_{DS}(d) = \sum R_n \cos(n \cos^{-1}(1 - \frac{d}{D_0}) + \phi_n)$, where $D_0$ is the oscillation amplitude of the tuning fork. To accurately capture the photocurrent unaffected by other influences, two sets of $I_{PC}$ data were acquired in each scan, one with the light on and the other with the light off. This procedure was essential to eliminate the floating gate effect caused by the tip-modulation scanning probe on the local electrical parameters of the device[27], an influence that can significantly exceed the magnitude of the photocurrent effect itself. In this discussion, we focus solely on presenting the difference of the two data sets, which represents the net photocurrent attributed solely to the superfocused light source. Figure 2b displays the harmonic components (include their signs) and reconstructed approach curves at four representative positions: the interior region of the contact (curve A), the electrode-side of the contact edge (curve B), the channel-side of the contact edge (curve C), and the FET channel region (curve D). Notably, all curves exhibit an inflection point between 10 to 20 nm, suggesting the presence of two distinct photocurrent mechanisms with different characteristic distances that contribute oppositely to the photocurrent generation.

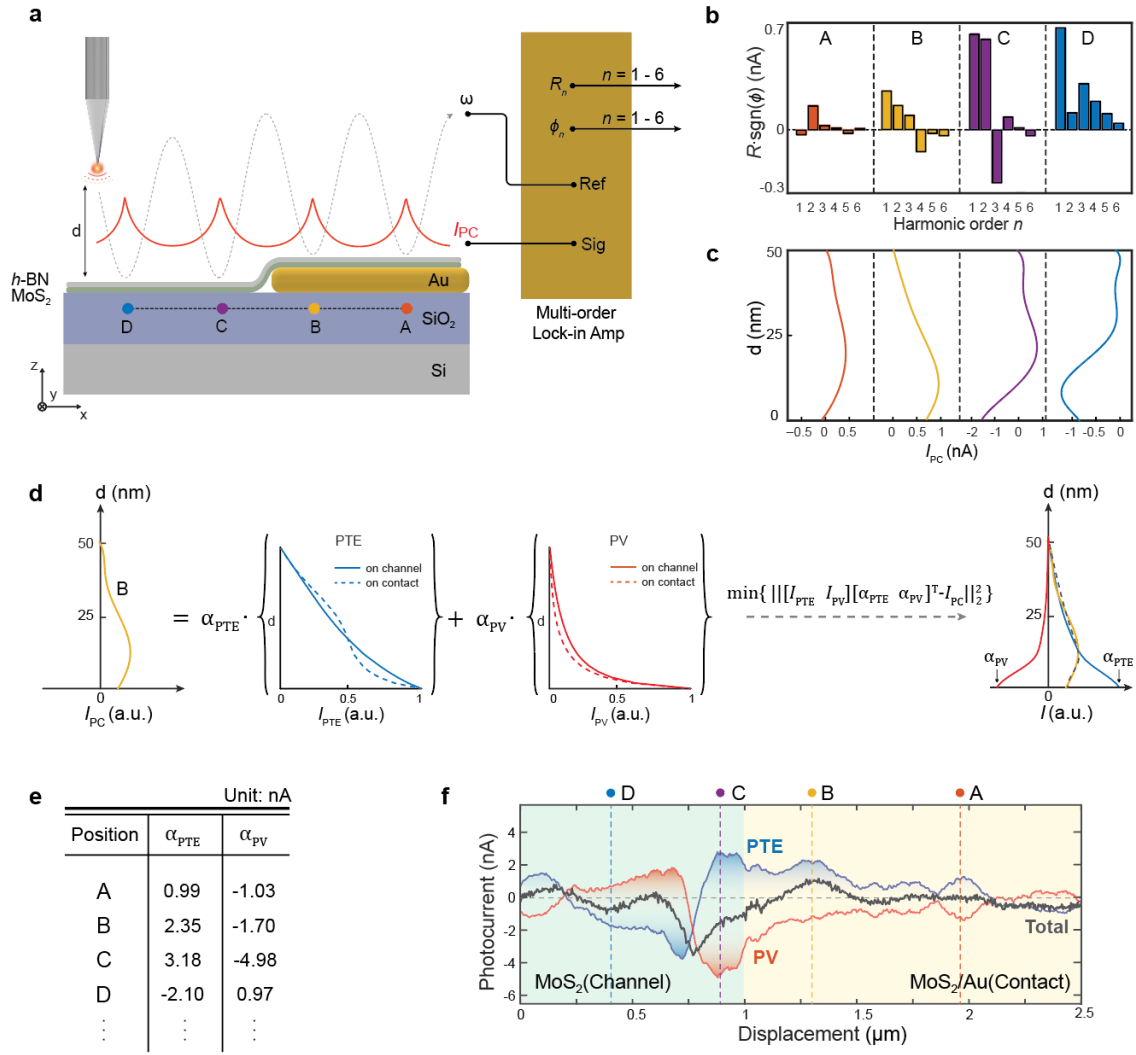

**Figure 2. Methodology and analysis of 3D near-field photocurrent microscopy**. **(a)** Experimental setup and operating principle. A superfocused light source (671 nm in wavelength, ~ 10 μW) illuminates the sample from the near-field, generating photocurrent signals for multi-order lock-in analysis. **(b)** Harmonic amplitudes and signs obtained from lock-in analysis at four representative locations on the sample. **(c)** Approach curves reconstructed using the data from panel (b). **(d)** Decomposition algorithm for analyzing PTE and PV contributions to the photocurrent. **(e)** Results of the decomposition analysis for data in panels (b) and (c). **(f)** Line profiles of total (gray), PTE (blue), and PV (red) photocurrents across the junction.

To differentiate these mechanisms, multiphysics simulations were used to derive their characterization curves. The PV effect's signal strength is determined by the area integral over the sample surface of the electric field intensity emanating from the nanoscale light source, which increases sharply at small distances *d*. Conversely, the PTE voltage, influenced by local temperature rises and their spatial distribution, is modeled by $V_{\text{PTE}} \propto \int T dx$, under the assumption of a linear spatial variation in the Seebeck coefficient (*S*) at the nanoscale. As both the magnitude and spread of the temperature in a hot spot contribute to this integral, the PTE effect shows reduced sensitivity to distance changes *d*. Additionally, the high thermal conductivity of the encapsulating hBN flake mitigates the temperature rise at smaller distances, further diminishing the PTE effect's sensitivity to changes in *d*. The approach curves for both PTE and PV effects, capturing these nuanced dynamics, are illustrated in Figure 2d.

To precisely quantify the contributions of PTE and PV to the total photocurrent $I_{PC}$, we employed a linear least-squares fitting approach. This involved optimizing coefficients, $\alpha_{PTE}$ and $\alpha_{PV}$, to minimize the residual sum of squares, $||[I_{PTE}\ I_{PV}][\alpha_{PTE}\ \alpha_{PV}]^T - I_{PC}||_2^2$, where $||\cdot||_2^2$ denotes the Euclidean norm. This methodology enabled us to accurately determine the respective contributions of the PTE and PV mechanisms at different locations. The calculated coefficients, $\alpha_{PTE}$ and $\alpha_{PV}$, for four representative locations and along a line scan across the 2D-3D SB photodiode are illustrated in Figures 2e and f. These coefficients vary rapidly across the electrode edge, and importantly, their signs and magnitudes are not correlated with any specific harmonic orders. This observation supports our previous discussion about the complex nature of photocurrent generation and the limitations of using single harmonic orders to directly interpret photocurrents.

To elucidate the photocurrent generation mechanisms in the SB device, we systematically examined their distributions by varying both the bias voltage ($V_{DS}$) and back gate voltage ($V_{GS}$), which modulates the band structure and carrier concentration in $MoS_2$. Specifically, we conducted repeated line scans across the junction under various conditions. The incident laser power was set at around 10 µW with a wavelength of 671 nm. The observations align with the energy band diagram of the n-type contact/$MoS_2$ interface model. As illustrated in Figure 3a, a combination of top and edge contacts characterizes the morphology of a 3D metal contact to a 2D

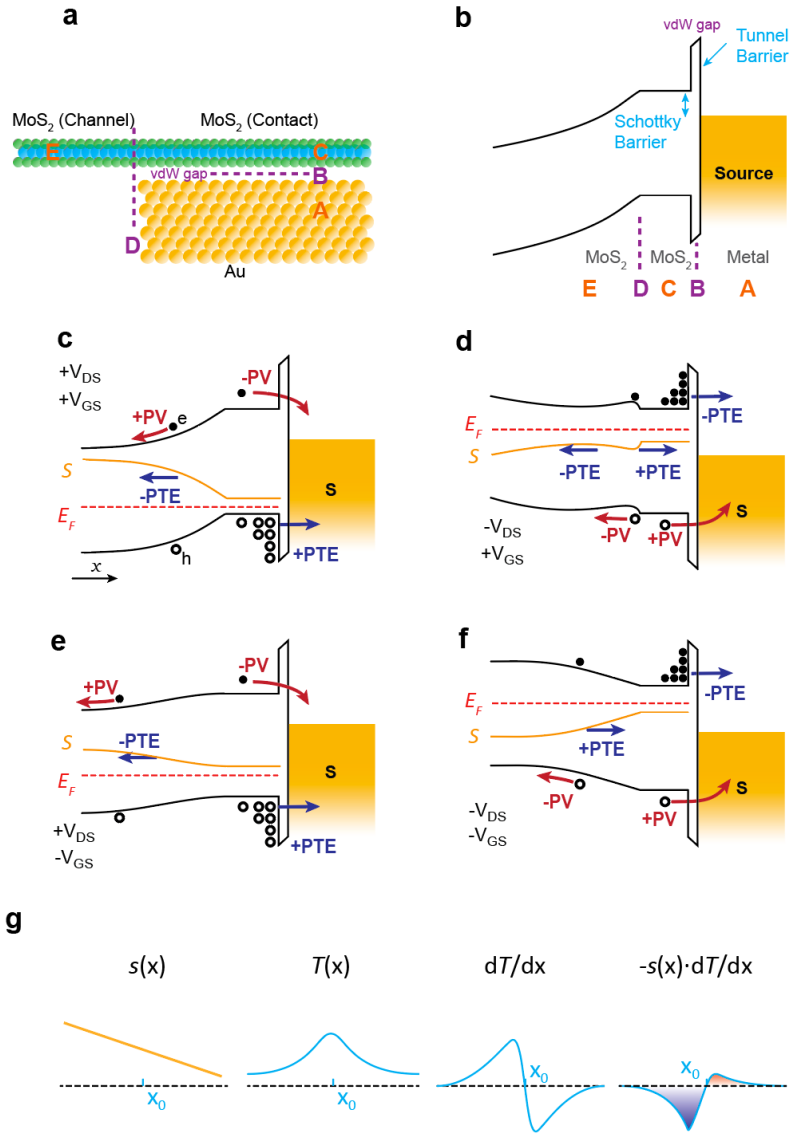

**Figure 3. Photocurrent response mechanisms under varied electrical biases and gate voltages.** (**a**) Cross-sectional illustration of the transferred MoS$_2$ on top of Au electrode. A, C, and E denote the three regions while B and D are the two interfaces separating them. (**b**) Energy band diagrams displaying the Schottky barrier and vdW gap influence under typical operating conditions. (**c-f**) Band alignment illustrations depicting the generation of PV and PTE currents at different regions, under different bias ($V_{DS}$) and gate voltage ($V_{GS}$) settings. **g**, Schematic representation PTE current generation at the depletion region, driven by spatial variations in the Seebeck coefficient $S$ and temperature gradient $dT/dx$ along the device channel.

MoS$_2$ film. The interface between metals and MoS$_2$ in a bottom-contacted configuration does not form covalent bonds but is instead established through a van der Waals (vdW) gap[28]. This vdW gap serves as an additional tunnel barrier for carriers, existing alongside the inherent Schottky barrier, and can hold either holes or electrons in MoS$_2$, depending on the polarity of the applied bias voltage $V_{DS}$.

The polarity of both PTE and PV currents demonstrates dynamic reversals at the electrode edge, contingent upon $V_{DS}$. With a positive $V_{DS}$, the PTE current is positive and the PV current is negative within the electrode contact region. This scenario occurs as holes accumulate at the MoS$_2$ side of the vdW gap, inducing a positive $S_{\mathrm{MoS}_2}$. The resulting difference between $S_{\mathrm{MoS}_2}$ and $S_{\mathrm{Au}}$ of the gold electrode (~1.5 µV/K[29]) generates a positive cross-interface PTE current, modeled by $(S_{\mathrm{MoS}_2} - S_{\mathrm{Au}})\Delta T$, where $\Delta T$ denotes the temperature difference across the interface. In contrast, the observed negative PV current primarily originates from the thermionic emission of photo-excited electrons transitioning from MoS$_2$ to the gold electrode, with holes hindered by a thin hole layer in MoS$_2$ above the MoS$_2$-Au interface. As depicted in Figure 4a, under positive $V_{DS}$, the PV current comprises roughly half the magnitude of the PTE current, resulting in a net positive current that corroborates findings from traditional scanning photocurrent microscopy techniques. Conversely, with negative $V_{DS}$ (illustrated in the bottom half of Figure 4a), the restrictive vdW gap impedes electron flow, leading to negative $S_{\mathrm{MoS}_2}$ and correspondingly negative PTE currents. Meanwhile, the blocked photo-excited electrons contribute to a positive PV current facilitated by hole diffusion, emphasizing the intricate interplay of charge carrier dynamics within the device.

For the first time, we report pronounced PTE and PV currents within the 2D depletion region of the MoS$_2$ channel, spanning several hundred nanometers from the interface of the 3D metal-2D semiconductor contact, under positive $V_{DS}$. The presence of a substantial PV current, as illustrated in Figure 3c, confirms significant band bending in the depletion region. This condition induces a rapid spatial variation of charge carriers, leading to a lateral inhomogeneity in the Seebeck coefficient $S_{\mathrm{MoS}_2}$, which follows the Mott relation derived from the energy dependence of the conductivity[30]:

$$S_{\mathrm{MoS}_2} = \frac{\pi^2 k_B^2 T}{3e} \frac{d\ln(\sigma(E))}{dE}\bigg|_{E=E_F}$$

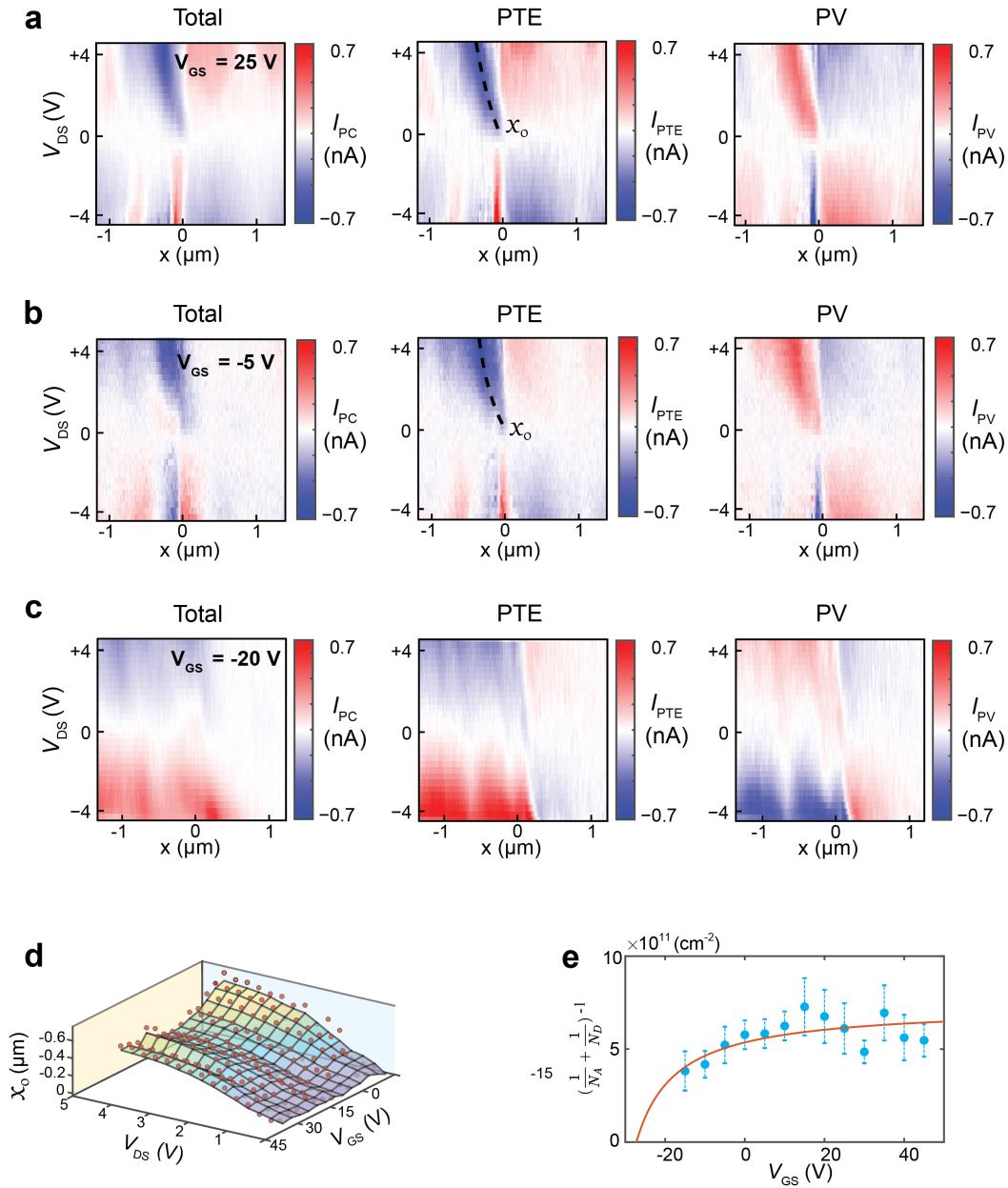

**Figure 4 Visualization of Photocurrent Generation Mechanism Under Varying Bias Conditions. (a-c)**, the total, PTE and PV currents at gate voltages of +25 V, -5 V, and -20 V, respectively, demonstrating the dependence of photocurrent mechanisms on electrostatic gating and biasing conditions. **(d)**, 3D plot correlating the depletion region width $x_0$ with changes in $V_{DS}$ and $V_{GS}$. **(e),** Variation of charge carrier concentration as a function of $V_{GS}$, with the red curve representing the fit based on a parallel plate capacitor model.

where $e$ is the electron charge, and $\sigma$ is the electrical conductivity, differentiated with respect to energy $E$ is evaluated at the Fermi energy $E_F$. When $E_F$ is positioned within the band gap, the Seebeck coefficient can be approximated by $S = \frac{E_F - E_b}{eT} + A$, where $E_b$ is the band edge energy and $A$ is a constant. The choice of $E_b$ as the valance or conduction band edge depends on the predominate type of carriers in the semiconductor. As depicted in Figure 3g, a monolithically decreasing $S_{\text{MoS}_2}$ at a thermal hotspot induces a PTE current, represented by $I_{PTE} = \int -S_{\text{MoS}_2} \frac{dT}{dx} dx$, resulting in an overall negative current. Conversely, a monotonically increasing $S_{\text{MoS}_2}$ yields a positive PTE current.

Under the combined influence of negative $V_{DS}$ and positive $V_{GS}$, the band bending at the MoS$_2$-source interface is relatively weak, as illustrated in Figure 3d. Consequently, both the PTE and PV currents are diminished. In this setup, the most significant spatial variation in $S_{\text{MoS}_2}$ occurs at the electrode edge—the starting point of the 2D depletion region, where band bending is most intense. This results in the observed narrow profiles of both PTE and PV currents under negative $V_{DS}$ in Figure 4a.

In contrast, when $V_{GS}$ is negative, as depicted in Figures 3e and f, the depletion of intrinsic electrons within MoS$_2$ leads to a notably expanded depletion region. This expansion causes a slow variation in band structures across a wider area, inducing substantial fluctuations in $S_{\text{MoS}_2}$ and enhancing the PTE response, detailed in Figure 4c. Additionally, due to the electron depletion under negative $V_{GS}$, MoS$_2$ shows increased resistivity, affecting the distribution of voltage drop across the MoS$_2$ channel and the Schottky barriers. This redistribution of bias potential intensifies the electric field in channel and the separation of photocarriers, thereby amplifying the PV effect. Despite the predominance of PTE, the total current under $V_{GS} = -20$ V is reduced compared to conditions with higher $V_{GS}$, underscoring a nuanced interplay between resistive effects and carrier dynamics that subtly modulates the overall photocurrent.

To enable a more precise quantitative comparison, we conducted the further examination of the measurements from the depletion region. At the edge of the contact, the depletion layer width is notably greater than in the bulk MoS$_2$. This is attributed to the relatively weaker screening in 2D lateral p-n junctions compared to their 3D counterparts, a phenomenon experimentally validated through optical characterization and Kelvin Probe Force Microscopy (KPFM)[31,32]. In the contact region, the classical 3D Schottky barrier model is applicable:

$$L|_{3D} = \sqrt{\frac{2\varepsilon_s V_s}{e}\left(\frac{1}{N_A} + \frac{1}{N_D}\right)}$$

where $\varepsilon_s$ represents the semiconductor permittivity, $V_s$ the built-in potential, and $N_A$ and $N_D$ the acceptor and donor concentrations, respectively. Utilizing $N_D = 2 \times 10^{18}\ cm^{-3}$ from IV characterization, a depletion width of a few nanometers is expected in the contact region. At the junction edge, the depletion length is significantly larger due to the weak electrostatic forces inherent in 2D materials. It is expressed as[33]:

$$L|_{3D\ metal-2D\ semi} = \frac{\pi^2 \varepsilon_{eff} V_s}{8Ge}\left(\frac{1}{N_A} + \frac{1}{N_D}\right)$$

Here, $\varepsilon_{eff}$ is the effective permittivity considering the surrounding dielectric, and $G$ is Catalan's constant ($G \approx 0.915$). The depletion length $L$ scales approximately linearly with the voltage applied across the interface. With a positive $V_{DS}$, the interface between MoS$_2$ and the source experiences an inverse bias, thereby dominating the distribution of the applied voltage, which closely aligns with $V_{DS}$. The depletion length $L$ can be substituted by $x_0$, denoting the largest slope region in the bending band. This point is identified from the locations of peak PTE current under varying $V_{DS}$ conditions, as depicted in Figure 4c. By linearly fitting $x_0$ against $V_{DS}$, we calculate the effective carrier concentration $\left(\frac{1}{N_A} + \frac{1}{N_D}\right)^{-1}$ for different $V_{GS}$ settings, detailed in Figure 4d. This fit aligns with the parallel plate capacitor model, revealing a charge-neutral point at $V_{CNP} = -27$ V, where the depletion length can extend indefinitely.

Since the PTE current observed in our studies is determined by the integral $\int S_{\text{MoS}_2} \frac{dT}{dx} dx$, it is influenced not only by the spatial variation of $S_{\text{MoS}_2}$ but also critically by the thermal profile of the hotspot. Effective thermal management at the nanoscale, therefore, becomes essential. Conventionally, thermoelectric effects are optimized in materials with low thermal conductivity, which helps maintain the necessary temperature gradient across the device to facilitate electrical power generation. However, in the case of hotspots, the scenario differs; both the intensity and the spatial distribution of the thermal profile are pivotal, jointly dictating the thermoelectric performance. For instance, when a thin MoS$_2$ film is placed on a glass substrate—as schematically shown in Figure 5a—the lower thermal conductivity of MoS$_2$ compared to glass leads to predominant vertical heat dissipation into the substrate rather than lateral spreading. This results in a hotspot with a full width at half-maximum (FWHM) of only 20 nm, as shown by the simulation results in Figure 5g. The introduction of a 10 nm layer of hBN modifies this dynamic significantly. The extraordinary high thermal conductivity in hBN facilitates lateral heat dispersion, expanding the FWHM of the thermal profile to approximately 150 nm. Such an expanded thermal gradient effectively captures the spatial variations in $S_{\text{MoS}_2}$, thereby substantially enhancing the PTE effect. This demonstrates the critical role of layering and material choices in the design of nanoscale thermoelectric devices, where the configuration can dramatically influence the efficiency of energy conversion.

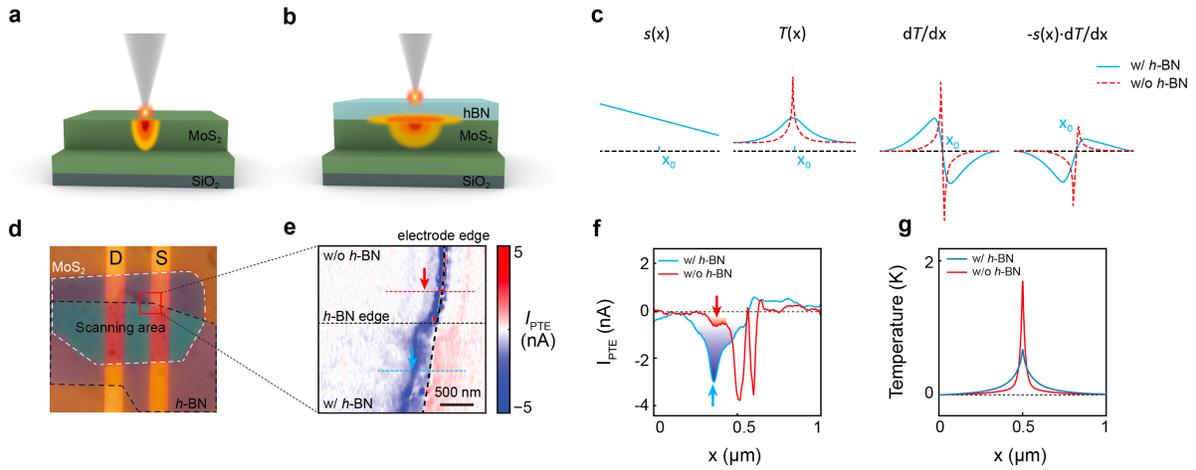

**Figure 4. Influence of hBN encapsulation on PTE current and thermal profiles in MoS2 Schottky Barriers. (a-b)** Schematic representation of the temperature profiles in MoS2 devices, with (b) and without (a) hBN encapsulation. **(c)** Comparison of the PTE current generated by different thermal profiles. **(d)** Optical microscope image of a device partially covered by a 10-nm-thick hBN layer. **(e)** PTE current image near the contact electrode, highlighting the enhancement from hBN layer. The blue and red arrows indicate the positions of $x_0$. **(f)** Line scans through different device regions demonstrating the comparative PTE amplitude variations. **(g)** Simulated temperature profiles within MoS2.

To substantiate the beneficial role of hBN on the PTE current, we conducted an experimental setup where a Schottky barrier photodetector was partially covered by a 10-nm-thick hBN flake. The optical image of this setup is presented in Figure 5d, and the corresponding PTE current measurements are depicted in Figure 5e. Remarkably, in the region encapsulated by hBN, the maximum PTE current near $x_0$—positioned approximately 300 nm from the electrode edge—was clearly observed. Conversely, in the bare MoS$_2$ area, the PTE current exhibited a significant decrease, dropping from about -3 nA to approximately -0.5 nA. This drastic reduction underscores the thermal management efficacy of the hBN layer, which enhances lateral heat dispersion and thereby amplifies the PTE response.

The unchanged position of $x_0$ in the bare MoS$_2$ region suggests that the absence of hBN does not significantly alter the depletion layer width within the device, which is expected since a 10-nm thick hBN layer does not substantially change the effective permittivity $\varepsilon_{eff}$ for MoS$_2$. However, a significant PTE current was notably observed at the electrode edge in the bare MoS$_2$ region, driven primarily by a sharp variation in $S_{\text{MoS}_2}$ confined to a narrow region around the electrode edge. Notably, this pronounced edge PTE current was exclusively detected under negative $V_{DS}$ conditions (as shown in Figure 3g). Such observations not only validate the thermal modulation effects of hBN but also demonstrate how material interfaces and coverings critically influence the distribution and intensity of PTE currents in nanoscale devices.

In summary, this study introduces a three-dimensional photocurrent nano-imaging technique to delineate the mechanisms of photocurrent generation within a Schottky photodetector with nanoscale spatial resolution. By leveraging reconstructed approach curves for each pixel, we have successfully distinguished various photocurrent generation mechanisms based on their unique signatures. Notably, we detected a pronounced photothermoelectric current that reaches far into the semiconductor material, extending several hundred nanometers from the electrode contact edge, a phenomenon enhanced by the weak electrostatic forces characteristic of 2D materials. Moreover, we demonstrated that integrating a thin layer of hBN, a material with high thermal conductivity, enhances the PTE response. This enhancement is achieved by broadening the temperature gradient region, thereby optimizing the spatial match with variations in the Seebeck coefficient. Our findings not only pave the way for the development of high-performance integrated photodetectors but also establish a novel methodology for probing light-matter interactions at the nanoscale.